\documentclass[12pt,aps,prb,preprint]{revtex4}   
\usepackage{amsmath}    
\usepackage{amssymb}
\usepackage{graphicx}   
\newcommand{\med}{$m_{1/2}$ }
\newcommand{\medm}{m_{1/2} }
\newcommand{\be}{\begin{equation}}
\newcommand{\ee}{\end{equation}}
\newcommand{\beq}{\begin{eqnarray*}}
\newcommand{\beqn}{\begin{eqnarray}}
\newcommand{\eeq}{\end{eqnarray*}}
\newcommand{\eeqn}{\end{eqnarray}}
\newcommand{\rv}{random variable }
\newcommand{\rvs}{random variables }

\newcommand{\dem}{\frac{1}{2}}
\begin{document}
\pagestyle{headings}
\title{Use of the median in physics and astronomy}
\author{Jean-Michel L\'evy}
 \altaffiliation[Also at ]{home.}  
 \affiliation{LPNHE, Universit\'e Pierre et Marie Curie, 4 place Jussieu, 75230 Paris Cedex05}
 \email{jmlevy@in2p3.fr}   
\date{\today}
\begin{abstract}
The sample median is often used in statistical analyses of physical or astronomical data 
wherein a central value must be found from samples polluted by elements which do not  
belong to the population of interest or when the underlying probability law is such that 
the sample mean is useless for the stated purpose.\\
 Although it does not generally possesses the nice linearity properties of the mean, 
the median has advantages of its own, some of which are explored in this paper which elucidates
analogies and differences between these two central value descriptors. Some elementary
results are shown, most of which are certainly not new but not widely known either.\\
It is observed that the moment and the quantile approaches to the description of a probability distribution 
are difficult to relate, but that when the quantile description is used, the sample median can be  
characterized very much in the same way as the sample mean; this opens the possibility of using it 
for estimation purposes beyond what is usually done.\\
In order to relate the two approaches, a derivation is given of the asymptotic joint distribution of 
the mean and the median for a general continuous probability law.\\
\end{abstract}
\maketitle
\hfill{}

\section{Introduction}

As a measure of the central value of a continuous probability distribution, the median which 
halves it into pieces of equal probabilities is certainly as valuable an indicator 
as the expectation value (E.V.). It has the further advantage to exist for all continuous 
distributions, contrary to the E.V. which is not always defined,
as examplified by the well known lorentzian (alias Breit-Wigner, alias Cauchy) density function.\\ 

In practical problems, the superiority of the sample median over the mean manifests itself when
a central value must be derived from samples polluted by data which do not belong to the population 
of interest. In thoses cases, it has the advantage of a lower sensitivity to 'outliers', that is, 
to abnormally high or low values which most likely come from the contaminating data.\\

It is shown in this paper that, given a sample of $n$ independent random variables with the same 
continuous parent distribution, the sample median possesses, with respect to the parent 
distribution median \cite{vocabl}, properties which are quite similar to those of the sample mean w.r.t. the 
parent E.V. It will be argued that the often invoked difficulties in the calculation 
of the median distribution characteristics arise from ill-posed questions, but that there exist 
simple answers provided one does not try to fit a round peg in a square hole. To take but one example, 
it is in general not possible to analytically calculate the E.V. of the sample median, but 
attempting to perform this calculation is trying to answer the wrong question: it is the distribution {\bf median} 
(and not the mean) of the sample median which ought to be calculated and this can be done easily and in full 
generality.
A by-product of this point of view is that the often made comparison between the merits of the 
sample median and the sample mean based on their respective standard deviations, which uses expectation values,
might not be as well grounded as it is thought to be and should not necessarily lead to prefer the mean.\\

This is not to say that there are no problems in using the median; first and foremost, 
the distribution median is generally not defined stricto sensu for a discrete random variable 
and the sample median for discrete data needs some kind 
of interpolation to be defined; we shall not deal with those problems here and content 
ourselves with discussing continuous distributions. But even for these, we must further restrict
ourselves to 1-dimensionnal random variables; applying the standard definition to a multidimensionnal
distribution would yield a median point which varies with the chosen axes.\\ 
Another drawback arises from the fact that the median is not, in general, a linear
operator over the vector space of random variables; the question of deciding when it is linear is
a difficult one (obviously akin to the problem of the multidimensionnal median just evoked)
and we shall not deal with it in full generality in this note.\\
The third kind of problem has been evoked above; it arises from ill-posed questions and we shall see how
to dispose of it. \\

The plan of this paper is as follows:\\
Part II is a very short reminder of basic facts about order statistics which will be needed in the 
sequel. \\
Part III is devoted to demonstrate that the median (we are speaking here of the random variable as well as of the 50\% 
quantile) can arguably be used to characterize the "center" of a distribution as legitimately as the mean (we are speaking 
here of the random variable as well as of the first moment); both random variables are shown to behave similarly 
w.r.t. to the samples from which they are derived. It is further remarked that the median possesses some qualities which 
have no counterpart for the expectation value. (obviously, the converse is also true) \\
Part IV makes use of these findings to show how the median can be employed for estimation purposes. The example of the
estimation of a ratio is explored.\\
Part V is devoted to a short attempt at defining a dispersion characteristic for the median in line with 
the quantile type of approach to a probability distribution. The variance, being an expectation 
value, is not well fit for this r\^ole and we propose other solutions based on intervals. \\
Part VI is slightly off the main line of argument, but is included for completness and because it 
might be of use to people who need to define the centroid of a distribution by some linear combination of the 
mean and the median.  The full asymptotic joint probability distribution of these two statistics is derived;
to the best of the author's knowledge, the expression of the covariance is new. \\
Part VII contains our conclusions. An appendix is added to demonstrate that the problem of building an unbiassed
ratio estimator addressed in part IV is not solvable in the usual sense.\\
For pedagogical purposes, all demonstrations are given in full, even when they can be easily found in the litterature.
 
\section{Order statistics and distribution quantiles}

{\it From now on and unless otherwise stated, it will be understood that the parent distribution of the 
random variables forming the samples that we shall deal with is continuous and that its c.d.f. 
(cumulative distribution function) is strictly increasing on its domain of variation. As a consequence, for any $q \in 
]0,1[$ there exists one and only one value $m_q$ such that $F(m_q) = q$. $m_q$ is known as the q-quantile of the 
distribution.
The particular case $q = 1/2$ corresponds to the (distribution-)median \med  \\

Given a sample of $n$ random variables $X_1, X_2 ..X_i..X_n$, we define its $k^{\rm th}$ order statistics or its 
$k^{\rm th}$ quantile $Y_k$ as the random variable which takes the value of the $k^{\rm th}$ of 
the $X_i's$ when these are renumbered in ascending order. In other words, $Y_1$ is defined as the smallest of 
$X_1..X_n$, $Y_2$ as the next smallest and so on, for any realisation of the sample. \\

In what follows, we shall assume that the $X's$ are i.i.d. (independent, identically distributed)  with continuous 
c.d.f. $F(x)$ and, would the need arise, p.d.f. $f(x) =\frac{dF}{dx}(x)$. Observe that, together with the hypothesis 
stated at the beginning of this section, this implies that there is zero probability that any two of the $X's$ will be 
equal. The $Y's$ are therefore unambiguously defined.} \\

{\bf Lemma 1~:} The distribution of the $k^{\rm th}$ order statistics is given by 
\be dF_k(x) = \frac{1}{B(k,n-k+1)}F(x)^{k-1}(1-F(x))^{n-k}dF(x) ~~~\cite{beta} \label{one} \ee

{\bf Proof~:} $Y_k$ lies between $x$ and $x+dx$ if and only if $k-1$ among the $X's$ take their values below $x$ 
(probability $F(x)^{k-1}$), one takes its value between $x$ and $x+dx$ (probability $\simeq dF(x)$) and the remaining 
($n-k$) 
take their values above $x+dx$ (probability $(1-F(x+dx))^{n-k}$). However, the first $k-1$ can be choosen in $C_n^{k-1}$
mutually exclusive ways to each of which can be associated $n-k+1$ exclusive ways of choosing the $k^{\rm th}$. 
Neglecting the $dx$ in $F(x+dx)$ which could only generate terms of second order and higher,  
we obtain the stated result.\\

{\bf Lemma 2~:} The joint distribution of the $k^{\rm th}$ and $l^{\rm th}$ order statistics is given by 
\be dF_{k,l}(x,y)\\=\frac{1}{B(k,l-k)B(l,n-l+1)}F(x)^{k-1}(F(y)-F(x))^{l-k-1}(1-F(y))^{n-l}dF(x)dF(y) \label{two} 
\ee for $x < y$ and $0$ for $x \geq y$ \\

{\bf Proof~:} $Y_k$ lies between $x$ and $x+dx$ and $Y_l$ between $y$ and $y+dy$ iff $k-1$ among the $X's$ take their 
values below $x$, (probability $F(x)^{k-1}$),  one takes its value between $x$ and $x+dx$ (probability $\simeq dF(x)$) 
$l-k-1$ among the remaining $n-k$ take their values between $x$ and $y$ (probability $(F(y)-F(x))^{l-k-1}$), one takes
its value between $y$ and $y+dy$ (prob. $\simeq dF(y)$) and the remaining $n-l$ are all above $y$ (prob. 
$(1-F(y)^{n-l}$). 
Counting the number of mutually exclusive choices for the indices entering those various sets and simplifying the result 
yields the given numerical coefficient.\\

Clearly, these results can be extented to an arbitrary number of sample quantiles but we shall not need more in this 
paper. Note however the obvious but interesting fact that after the change of variable $x \rightarrow F(x)$ those 
distributions are totally independent of the parent probability law. The $F(Y_k)$'s follow $\beta_I$-type probability 
distributions and multidimensional extensions thereof. This is very usefull when constructing confidence 
intervals for the distribution quantiles.

\section{The case of the median}
As already remarked, the hypotheses entail that there exists a unique value \med of the argument 
of $F$ for which $F({\medm}) =1/2$ . \med \cite{note1}
is then known as the parent {\bf distribution median} which we shall presently relate to the {\bf sample median} 
defined as the central value for an odd-sized sample ($r^{\rm th}$-order statistics if $n=2r+1$) or whichever of the two 
central values ($r^{\rm th}$ or $(+1^{\rm th}$-order statistics) comes up in an even odds random draw for an 
even-sized sample ($n=2r$).\\

{\bf Lemma 3~:} for an odd-sized sample of size $2r+1$, the probability that the median $M$ lies 
between $x$ and $x+dx$ is $(r+1)C_{2r+1}^rF(x)^r(1-F(x))^rdF(x)$ up to higher order terms. \\

{\bf Proof:}~ This is lemma 1 applied to $n = 2r + 1$ and $k = r + 1$ \\

The p.d.f. of the sample median is therefore:
$$\displaystyle f_M(x) = \frac{2r+1!}{(r!)^2}F(x)^r(1-F(x))^rf(x)$$

By similar use of lemma 1, it is immediate to show that for an even-sized sample of size $2r+2$, the density
of the $r+1^{\rm th}$ variable is:

$$\displaystyle f_{M1}(x) = \frac{2r+2!}{(r!)(r+1!)}F(x)^{r+1}(1-F(x))^rf(x)$$
and that of the  $r+2^{\rm th}$ variable is:
$$\displaystyle f_{M2}(x) = \frac{2r+2!}{(r!)(r+1!)}F(x)^{r}(1-F(x))^{r+1}f(x)$$
which, by averaging (following our definition of the median for an even sample)
gives us back the same distribution that we have found for the $2r+1$-sized sample.\\
This being established, we won't have to worry about the parity of the sample size in
most cases. 

\subsection{Properties shared with the mean}

In what follows, $M$ is assumed to be the median of an odd-sized sample of 
iid random variables of p.d.f. $f(x)$, and ${\cal M}$
stands for the "median operator", that is ${\cal M}[X]$ is the median of the
distribution of the random variable $X$ exactly as $E[X]$ is the expectation value of 
this distribution (remember that we restrict ourselves to distributions having a {\bf unique} median). 
$f_M(x)$ will denote the p.d.f. of $M$.\\

{\bf Theorem 1:~} the median of the distribution $f_M(x)$ is the same as that of $f(x)$,
that is ${\cal M}[M] = {\cal M}[X]$\\

{\bf Proof:~}
$\int_{-\infty}^uf_M(x)dx = \int_{0}^{F(u)}\frac{2r+1!}{(r!)^2}v^r(1-v)^rdv$ 
through the change of variable $v = F(x)$\\
However, the last integrand is form-invariant under $v \rightarrow 1-v$ and the integral 
up to $F=1$ must be equal to 1; therefore:
$$\int_{0}^{1/2}\frac{2r+1!}{(r!)^2}v^r(1-v)^rdv = 
\int_{1/2}^{1}\frac{2r+1!}{(r!)^2}v^r(1-v)^rdv = 1/2$$ 
Since $F(\medm) = 1/2$, it follows that $m_{1/2}$ is also the median of the distribution 
$f_M$\\

{\bf Remark 1:~} This is the perfect analogue of the equality between the E.V. 
of the sample mean and that of the parent distribution. Clearly, 
the E.V. of $M$ has nothing to do here.\\

{\bf Remark 2:~} This theorem solves the problem of a distribution-free point estimation of the median, contrary 
to what is stated in \cite{KS1}, but it must be remembered that the estimator is {\bf median unbiassed}.
Searching an {\bf expectation value unbiassed} estimator is not logically well grounded and it is clear that 
the E.V. of $M$ will generally be dependant on the particular $F$ at stake and difficult to calculate, with 
the obvious exception of symmetrical distributions (see below)
The question of interval estimation will be taken up later on.\\

{\bf Theorem 2:~} The median of a probability distribution is the number with respect to which the mean absolute 
deviation is minimal: \\

$\int_{\mathbb R}|x-\xi|f(x)dx$ is minimal for $\xi = \medm$\\

Write  $$h(\xi) \equiv \int_{\mathbb R}|x-\xi|f(x)dx = \int_{x < \xi}(\xi-x)f(x)dx +  \int_{x > \xi}(x-\xi)f(x)dx$$\\

Then $$\frac{dh}{d\xi}(\xi) = \int_{x < \xi}f(x)dx  - \int_{x > \xi}f(x)dx \;\;\;\mbox{\rm and}\;\;\; 
\frac{d^2h}{d\xi^2}(\xi) = 
2f(\xi) > 0$$
therefore the point $\xi=\medm$ is the only zero of $\frac{dh}{d\xi}$ and corresponds to a minimum of $h$ \\

This parallels the property of the E.V. to minimize the mean squared deviation. 
($f(a) = E[(X-a)^2]$ is minimal for $a=E[X]$ )\\

{\bf Theorem 3:~} The median of an odd-sized set of real numbers $x_j, \{j = 1..2r+1\}$ is the value which 
minimizes $f(\xi) = \sum_j|x_j - \xi|$ with respect to $\xi$\\

As before, write $f(\xi) = \sum_j|x_j - \xi| = \sum_{x_j < \xi}(\xi - x_j) + \sum_{x_j > \xi}(x_j-\xi)$\\

This (continuous) function of $\xi$ can be derived everywhere except at the points $x_j$: \\
$f'(\xi) = cardinal\{j|x_j < \xi\} - cardinal\{j|x_j > \xi\}$\\
$f'$ is discontinuous and piecewise constant, but clearly monotonous, negative for small $\xi$, 
positive for large $\xi$ and zero for $\xi = x_{r+1}$\\

Note that for an even-sized set of numbers, any value in the central interval is a solution and a legitimate median.\\

Here also we have a result quite analogous to the one which holds for the mean w.r.t. the sum 
of the squared deviations. ($f(a) = \frac{\sum(x_i-a)^2}{n}$ is minimal for $a=\frac{\sum x_i}{n}$) \\

{\bf Corollary:~} The sample median is the maximum likelihood estimator of the parameter $a$ (which 
equals the E.V. and the median) of "Laplace's first law of errors" the density of which reads: 
$$\frac{1}{2b}e^{-\frac{|x-a|}{b}}$$
Maximizing the log-likelihood w.r.t. $a$ is the same as minimizing $f(\xi)$ in theorem 3.\\ 
Once again, this is the analogue of the result for the sample mean of an iid gaussian sample.\\

{\bf Remark 3:~} The analogy between the sample median and the sample mean 
is apparently dimmed by the fact that the sample mean dispersion is easily calculated in terms of
the parent distribution variance, but that the same calculation with $f_M$ is obviously much 
more difficult. The point is that the dispersion also is expressed in terms of expectation
values and one is again confronted with the problem of asking the right question.  
One should therefore find a measure of dispersion which could be expressed in a way
closer to the median kind of philosophy. This could be the width of an interval containing 
the sample median with a given probability. The problem will be briefly addressed in part V.\\
Another point is that the E.V. is a linear operator on the vector space of
random variables which admit an E.V. Likewise, the arithmetic mean of sequences of length $n$ 
is a linear operator on the vector space ${\Bbb R}^n$. Such is generally not the case, neither for 
the distribution median, nor for the median of a numerical sequence. \cite{oldfoot}\\

However, the following two theorems which sum up intuitively obvious properties, show that 
for symmetrically distributed random variables the median operator is linear:\\

{\bf Theorem 4:~} If $f$ is symmetrical about $a$ then $\medm =a$ and $f_M$ is also symmetrical
about $a$. In addition, if $f$ has an expectation value $m$, then $\medm = m = E[M] = {\cal M}[M]$

{\bf Proof:~} Assume that $f(2a-x) = f(x)$.\\ Then $\int_{-\infty}^a f(x)dx = \int_{-\infty}^a
f(2a-x)dx = \int^{\infty}_a f(y)dy$ with $x \rightarrow 2a-y$ \\Since $\int_{\mathbb R} f(x)dx = 1$, 
one thus finds $\int_{-\infty}^a f(x)dx = 1/2$ and $a = \medm$. \\
Moreover $F(2a-x) = \int_{-\infty}^{2a-x}f(t)dt = \int_x^{\infty}f(2a-u)du = 1-F(x)$ using 
again $t \rightarrow 2a-u$ and the symmetry of $f$.\\
Thus $f_M(2a-x) = \frac{2r+1!}{(r!)^2}F(2a-x)^r(1-F(2a-x))^rf(2a-x) =  f_M(x)$\\
Now $m = \int_{\mathbb R} xf(x)dx =
\int_{\mathbb R} xf(2a -x)dx = \int_{\mathbb R} (2a-y)f(y)dy = 2a - m$\\
Therefore $a = m = \medm$ and since $F(1-F) \leq 1/4$ it is obvious that if the parent has an expectation value, 
so does $M$. Hence $a = E[M] = {\cal M}[M]$\\

{\bf Remark 4:~}  there is evidently no reason for $E[M]$ to be equal to $E[X]$ in the general case,
not even asymptotically (see below).\\

{\bf Theorem 5:~} Let $X$ and $Y$ be two independnt random variables the densities of which are 
symmetrical about $a$ and $b$ respectively. Then $Z = \alpha X + \beta Y$ is symmetrical 
about $\alpha a + \beta b$.

{\bf Proof:~} $f_Z(z) = \int_{\mathbb R} f_X(t)f_Y(z-t)dt = \int_{\mathbb R} f_X(ru + s) f_Y(z - ru - s)|r|du$
for any real numbers $r \neq 0$ and $ s$.\\
Now  $f_Z(2a+2b-z) = \int_{\mathbb R} f_X(r u + s)f_Y(2a+2b-z-ru-s)|r|du$ \\
but since $f_X(2a-x) = f_X(x)$ and $f_Y(2b-x) = f_Y(x)$ , this can be rewritten:\\
$f_Z(2a+2b-z) = \int_{\mathbb R} f_X(2a-r u - s)f_Y(z+ru+s-2a)|r|du$ \\
which, with $s = 2a$, $r = -1$,  is identical to the first expression for $f_Z(z)$\\

With the same hypotheses, ${\cal M}[X] = a$, ${\cal M}[Y] = b$ by theorem 4 and the last
result shows that ${\cal M}[\alpha X + \beta Y] = \alpha a + \beta b$. \\

{\bf Therefore $\cal M$ is a linear operator on the vector space of symmetricaly distributed random variables.}\\

Let us stress again that these linearity properties do not generally apply for non-symmetrical random 
variables. However the following is obvious:

{\bf Theorem 6:~}  Let $X$ and $Y$ be two continuous, independent, identically distributed random 
variables. Then the median of $X-Y$ is $0$\\

{\bf Proof:~} $f_{X-Y}(z) = \int_{\mathbb R} f(z+t)f(t)dt$ hence $f_{X-Y}(-z) = \int_{\mathbb R} f(-z+t)f(t)dt$
which can be rewritten $\int_{\mathbb R} f(t)f(t+z)dt $ by changing the integration variable $t \rightarrow z+t$\\
One can obviously also use a symmetry argument: $P(X < Y) = P(Y < X)$ \\

A barely less simplistic symmetry argument will be used to show the following:\\

{\bf Theorem 7:~} Let $X_i, \{i=1..2r+1\}$ be a set of independent \rvs having the same median \med 
but otherwise arbitrary distributions and let $M$ be defined as for an iid sample. Then ${\cal M}[M] = \medm$ .\\

This, again, is a property shared with the mean and the E.V. in which case it is a straightforward consequence
of linearity. For the median, however, another argument is required.\\

First note that without loss of generality, \med can be taken equal to $0$.\\

The hypothesis then reads: $P[X_i < 0] = 1/2 \;\;\forall i$  and if $M$ is the median of the $X_i$'s, one wants
to show that $P[M < 0] = 1/2$\\

Now $ M < 0$ if and only if at least $r+1$ among the $2r+1$ $\{X_i < 0\}$ events occur. However the number of these
is a binomial \rv ${\cal B}(2r+1,1/2)$ and the probability for having $k$ events exactly is  
$\frac{1}{2^{2r+1}}C_{2r+1}^k$\\ Therefore 
$P[M < 0] = \frac{1}{2^{2r+1}}\sum_{k \ge r+1} C_{2r+1}^k$. The sum on the right-hand side is easily seen
to equal half the total and the theorem is proved.

\subsection{Properties not shared with the mean} \label{invariance}
It is well known that quite generally $g(E[X]) \neq E[g(X)]$ except when $g$ is an affine
function of its argument. \\For example, if  $g$ is convex, Jensen's 
inequality holds: $ g(E[X]) < E[g(X)]$ \\In statistical inference this has the consequence that, 
if an unbiassed estimator can be found for some parameter of a probability distribution, 
no function (except affine) of this parameter can be estimated without bias using the same 
estimator.\\
From that point of view, the distribution median is much easier to manipulate. Since it is 
defined by $P(X < \medm) = 1/2$ it is immediate that for any continuous strictly monotonous 
$g$, $g(\medm)$ is the median of the distribution of $g(X)$. Indeed, $X < \medm$ is equivalent
to either $g(X) < g(\medm)$ or $g(X) > g(\medm)$. Hence $P(g(X) < g(\medm)) = 1/2$ in both cases.
We shall call this the {\bf "invariance"} property of the median in the sequel.
Applying it to an affine transformation, we get:
$$ {\cal M}[\alpha X + \beta] = \alpha {\cal M} + \beta $$ 
for any two constants $\alpha$ and $\beta$ without any symmetry hypothesis here.\\

Of course, this leaves us still far from the linearity of the E.V. operator, but 
the invariance property strengthens the case for an estimation theory wherein 'unbiasedness' 
is no longer defined w.r.t. the E.V. of the estimator, but w.r.t. its median. 
The rationale for such a definition is neither worth nor better than that for the usual definition. 
Clearly, there is nothing sacred about 'unbiassedness' being defined in terms of expectation value.\\

On the other hand, finding the distribution median of a function of several random variables
is not an easy task except in the trivial and uninteresting gaussian case since it always 
boils down to invert some sort of c.d.f. defined by a convolution.\\

\section{Using the median for estimation}

\subsection{Estimation of the distribution median}

Theorem 1 solves the problem of the point estimation of the distribution median (See remark 2):
The sample median $M$ as defined here both for even and odd sized samples is a (median) unbiassed 
estimator of the distribution median \med . \\

By the same token, it shows how median unbiassed
estimators can be constructed in principle. If a transformation of the parent \rv can be found 
such that its median is equal to the parameter to be estimated, the median of the transformed sample becomes a 
median unbiassed estimator of the said parameter.\\

A simple example could be the estimation of the unknown parameter $a>0$ of the power p.d.f.\\
$f(x) = ax^{a-1} \;{\rm if\;} x \in ]0,1] \;\;$ and $0$ otherwise. The distribution median is $2^{-1/a}$ and
therefore the median of the transformed sample $X_i \rightarrow -Log(2)/Log(X_i)$ is a median unbiassed estimator
of $a$. Of course there are more classical solutions and we do not claim that this one is the "best".\\

A slightly different formulation is the following: if the parameter of interest can be expressed as a monotonous  
function of the distribution median, then (by the invariance property) the same function calculated with the sample 
median yields a median-unbiassed estimator of the parameter.\\

For example, estimating the parameter of the exponential density $f_a(x) = ae^{-ax}\theta(x)$\cite{Heavi} can be done by
observing that $\medm = \frac{Log(2)}{a}$ and therefore $A = \frac{Log(2)}{M}$ with $M$ the sample median, is 
a median-unbiassed estimator of $a$.  Again, there are more classical solutions to which this one should be compared.\\

\subsection{Interval estimation}

Building a confidence interval for \med is also simple, but can be done in full rigour and generality 
only for certain values of the confidence level which depend on the sample size. \\
Indeed, lemma 2 shows that the random interval formed by any pair of sample quantiles contains a 
given distribution quantile with a calculable probability, independent of the parent distribution:
\begin{equation}
\begin{array}{lcr}
 P(Y_k < m_q < Y_l) =  P(F(Y_k) < q < F(Y_l)) = \int_0^q \int_q^1 dF_{k,l}(u,v) && \\
     = \int_0^q \int_q^1 \frac{1}{B(k,l-k)B(l,n-l+1)}u^{k-1}(v-u)^{l-k-1}(1-v)^{n-l}dudv 
&&
\end{array}\label{probint}
\end{equation}
where we have changed variables $x \rightarrow u=F(x)$ and $y \rightarrow v=F(y)$ in (\ref{two}).

This latter expression is evidently independent of $F$ and in the case of the median ($q= 1/2$), it seems 
reasonable to take $l = n-k+1$ so that both ends of the interval are determined by statistics playing symmetrical roles. 
\\

Expression (\ref{probint}) can be simplified. The demonstration given in \cite{KS2} is both unduly complicated and 
wrong. 
Here is the simple way:\\
$P(F(Y_k) < q) = P(F(Y_k) < q, F(Y_l) < q) + P(F(Y_k) < q, F(Y_l) > q)$ but since $Y_k < Y_l$ by definition, the
first term reduces to $P(F(Y_l) < q)$, therefore \be P(Y_k < m_q < Y_l) = P(F(Y_k) < q) - P(F(Y_l) < q) \label{fin}\ee
which is readily calculated in terms of incomplete $\beta$ functions according to lemma 1.  \\

{\bf Remark~6:} it is sometimes usefull to bear in mind the connection between probabilities such as $P(F(Y_k) < q)$ and 
the binomial distribution: indeed, $Y_k < m_q$ if and only if at least $k$ among the $X_i$ are below $m_q$ or equivalentely,
at least $k$ among the $F(X_i)$ are below $q$. But the $F(X_i)$ are independent and uniformly distributed betwee $0$ and 
$1$, and therefore the number of them which take their value below $q$ follows a binomial distribution of parameters $n$
and $q$. Hence $P(F(Y_k) < q)$ is the probability that a variate ${\cal B}(n,q)$ following the said binomial 
distribution takes a value $ \geq k$. 
On the other hand, the c.d.f. of a binomial can readily be written as an incomplete, normalized $\beta$ 
integral by deriving the sum of the individual probabilities w.r.t. $q$, simplifying the result and re-integrating 
w.r.t. $q$ with the appropriate end-point condition. This yields \\$P(F(Y_k) < q) = P({\cal B}(n,q) \ge k) = \int_0^q 
\frac{x^{k-1}(1-x)^{n-k}}{B(k,n-k+1)} dx$ in accordance with a straightforward application of lemma 1.\\
Thus, in terms of binomial probabilities, (\ref{fin}) can be rewritten \be P(Y_k < m_q < Y_l) = P(l > {\cal B}(n,q) \ge 
k)\ee  and for the median ($q = 1/2$), \be P(Y_k < m_{1/2} < Y_l) = (\frac{1}{2})^n\sum_{j=k}^{j=l-1} C_n^j 
 = (\frac{1}{2})^n\sum_{j=k}^{j=n-k} C_n^j \ee
where the symmetric choice $l=n-k+1$ has been made in the last expression.\\
 
{\bf Theorem 8:~} Let $X$ and $Y$ be two positive, continuous, independent and equally 
distributed random variables. Then the median of $Y/X$ is equal to $1$\\
{\bf Proof:~}
This theorem is equivalent to theorem 6 restricted to strictly positive random variables.
Another proof is as follows~:\\
Let $F$ and $f$ be the common c.d.f. and p.d.f of $X$ and $Y$. Because the two are independent, 
it follows that: $$P(\frac{Y}{X} < 1) = P(Y < X) = \int_0^{\infty} P(Y < z|X=z)f(z)dz = \int_0^{\infty} 
F(z)f(z)dz = 1/2$$
{\bf Corollary:~} If the last hypothesis is relaxed to "if $Y$ is distributed as $aX$ with 
$a$ a positive constant", then the median of $Y/X$ is equal to $a$.

\subsection{An example of application}
A recurrent problem is that of evaluating without 'bias' the ratio of the responses of two
detectors to the same signal. This arises, for example, in photometry where one would like, to
state things simply, assess the ratio of the responses of two telescopes to the light received 
from the same stars. The estimate should, of course, be based on two sets
of observations performed with both instruments on the same objects. \\

The standard way of treating this kind of problem is to assume that both measurements (of the same
object) have 'true values' $X_{true}$ augmented by random errors of zero E.V.
The ratio $\alpha$ of the 'true values' supposed to be constant is by definition the photometric ratio 
at stake. The question does not possess any unbiassed general answer in the usual sense (see appendix). 
The standard least squares estimator is not even consistent, its bias calculated to lowest order involving a term that 
does not go to zero in the large sample limit.\cite{noreg}  The best that we have found is to use a constrained least 
square to refit the two sets of values whilst imposing a proportionnality relation to the refitted values. The estimator 
of the ratio turns out to be the root of a highly nonlinear equation which can be simplified if one assumes that there 
is also a constant ratio between the standard deviations of the (random) errors. In this case, the non-linear equation 
reduces to a quadratic polynomial, allowing for an easy evaluation of bias and variance to lowest order in the error moments. 
In particular, the bias is found to go to zero as $\frac{1}{n}$.\cite{jml}\\   

One might, however, make different assumptions. For example, it could be supposed that the ratios of the 
paired signals from the same objects follow the same probability law and one could try to evaluate the median of this 
law which would be defined as the required ratio. According to theorem 1, the sample median is a (median)-unbiassed 
estimator of this parameter. \\
This would be the situation if {\bf relative errors} in each serie were identically distributed, in which case 
the measurements could be written $X_{i,true}(1+\delta_i)$ and $\alpha X_{i,true}(1+\eta_i)$ with identically distributed 
$\delta_i$ and $\eta_i$. One would be precisely in the situation suggested here, with the median of the ratio 
probability distribution equal to $\alpha{\cal M}[\frac{1+\eta}{1+\delta}].$\\
 If one further assumes that the distribution of the $\eta's$ is identical to that of the $\delta's$, the correction factor 
to $\alpha$ is equal to 1 according to theorem 7 and its corollary. In any case, the value which should really be used in 
predicting, say, the luminosity of the image of a new object through the second telescope given its measured luminosity
through the first, is the corrected $\alpha$ given here and estimated by the sample median, for the relevant quantity 
in such a comparison is the ratio of the measured quantities rather than the ratio of inaccessible and poorly defined 
'true values' \\
Note that using the mean of the ratios (instead of their median), would yield \underline{one of the worst estimators that
can be thought of.} It is easy to see that it is not even consistent, as is always the case when one tries to estimate
a (non linear) function value by the average of the function values instead of averaging first the (measured) 
would-be arguments and using their averages as arguments of the function. \\

One could also imagine that the measurements read $X_{i,true} + \delta_i$ and $\alpha X_{i,true} + \eta_i$ with the
{\bf absolute error} $\eta_i$ being distributed as $\alpha \delta_i$ Then the corollary of theorem 8 applies and all the
ratios have equal distribution median, which, through theorem 7, can be estimated by their sample median.
But in this case one also has a constant ratio between the variances which allows to simplify the least-square
fit as explained at the beginning of this subsection.  Moreover, finding a confidence interval in the present case
where the various ratios have the same median but otherwise different distributions would not be an easy task.\\

There are, of course, other choices using, for example, the ratio of the means (once 
again, {\bf not} the mean of the ratios ! ) But this might arguably be thought of as less effective since it does not
take into account the correlation between the responses of the two instruments to the signals from a given object. The 
same can be said of other estimators which agglomerate the numerators and denominators separately before combining them, 
as for example the geometric mean.

\section{Evaluating the dispersion of $M$}
As already observed, the variance being defined in terms of moments is not taylored for
an easy evaluation of the dispersion of a value which is characterized by probability conditions.
A more adequate definition must be based on the probability content of the interval assumed to
represent this dispersion. 
A possibility would be to take it symmetrical on the $F(x)$ scale.
If $\int_{q}^{1-q} \frac{2r+1!}{(r!)^2}v^r(1-v)^rdv = 68\%$ or equivalently $\int_0^q 
\frac{2r+1!}{(r!)^2}v^r(1-v)^rdv = 16\%$, one could take $F^{-1}(1-q) - F^{-1}(q)$ as a measure 
of the dispersion of $M$. \cite{note2}\\

 
Another solution could use the full width at half maximum 
$(FWHM)$ of $f_M$ written in terms of the $v$ variable and convert it back to the $x$ scale. 
In terms of $F$ (or $v$) the condition $v^r(1-v)^r = \dem (\frac{1}{4})^r$ yields immediately the
bounds $1/2 \pm \dem\sqrt{1-(1/2)^{1/r}}$  and one sees that, to lowest order in $1/r$,
the interval width on the $F$-scale is $\sqrt{\frac{\log 2}{r}}$ or $~\sim \sqrt{\frac{\log 
2}{r}}\frac{1}{f({\medm})}$ on the $x$-scale. As expected, this goes to zero as $1/\sqrt{r}$.
We shall see in the next paragraph that the asymptotic standard deviation is in fact $\frac{1}{2\sqrt{2r+1}f(\medm)}$
which shows that the relation between $\sigma_{asympt}$ and $FWHM_{approx}$ is the same as that of a gaussian curve.
The probability for $M$ to fall in this interval can easily be computed numerically. It decreases with $r$ but remains
above $.761$ up to $r$ values of several thousands. 
\section{Asymptotics}
The derivation of the quantiles asymptotic distributions can be found everywhere \cite{KS3} and 
that of the median is but a particular case. We do here something which we haven't found
in the litterature and which might be of some use to those who need a definition of the 'centroid'
of a distribution in terms of median and mean. We calculate the joint asymptotic 
distribution of the median and the mean for an independent $n=2r+1$-sample \\

\subsection{Derivation}
In this paragraph, the distribution median is written $\mu$ instead of \med for reasons which should become clear 
during the reading !\\

Let $f(x)$ be the parent distribution p.d.f. We assume that it has an E.V. $m$, a variance $\sigma^2$.
and a unique median $\mu$.\\
 
The joint sample p.d.f. is $\prod_{j=1}^{j=n}f(x_j)$, but for the reordered sample $\{Y_j\}$, this becomes 
$n!\prod_{j=1}^{j=n}f(y_j)$ on the domain $D = \{y_1 \le y_2 \le y_3 ..\le y_n\}$\\
The sample mean $\bar{X}=\bar{Y}$ is known to have an asymptotically degenerate distribution \\$\delta(x-m)$ and it is $\sqrt{n}(\bar{X}-m)$
which converges to a gaussian ${\cal N}(0,\sigma^2)$.
What we have found about the dispersion of ${ M} = Y_{r+1}$ points to a similar behaviour and 
we therefore define the two random variables $L = \sqrt{n}(\bar{Y}-m)$ and $H = 
\sqrt{n}(Y_{r+1}-\mu)$ and calculate their joint 
p.d.f. according to:
$$g(l,h) = \int_D n!\prod_{j=1}^{j=n}dy_jf(y_j) 
\delta(l-\frac{1}{\sqrt{n}}\sum_j y_j 
+\sqrt{n}m))\delta(h-\sqrt{n}y_{r+1}+\sqrt{n}\mu)$$
Since the order between the first $r$ and the last $r$ integration variables is irrelevant, this can be simplified to:
$$g(l,h) = \int_{D'} \frac{n!}{(r!)^2}\prod_{j=1}^{j=n}dy_jf(y_j) 
\delta(l-\frac{1}{\sqrt{n}}\sum_j y_j + \sqrt{n}m))\delta(h-\sqrt{n}y_{r+1}+\sqrt{n}\mu)$$

With $D' = \{y_1,y_2,..y_r < y_{r+1} < y_{r+2}, y_{r+3},..y_n\}$\\

The $y_{r+1}$ integration amounts to a simple change of variable:
$$g(l,h) = 
\frac{n!}{(r!)^2\sqrt{n}}f(\mu+\frac{h}{\sqrt{n}})\int_{D'}\prod_{j=1}^{j=r}\prod_{j=r+2}^{j=n}dy_jf(y_j)
\delta(l-\frac{h}{n} -\frac{\mu}{\sqrt{n}}-\frac{1}{\sqrt{n}}\sum_{j\neq r+1} y_j + 
\sqrt{n}m)$$
In order to perform the $y$-integrations, we now use the Fourier representation of Dirac's 
$\delta$:
\beq g(l,h) = &&\int \frac{dt}{2\pi} \int_{D'}
\frac{n!}{(r!)^2\sqrt{n}}f(\mu+\frac{h}{\sqrt{n}}) 
\int_{D'} \prod_{j=1}^{j=r}\prod_{j=r+2}^{j=n}dy_jf(y_j)
e^{it(l-\frac{h}{n} +\frac{m-\mu}{\sqrt{n}}-\frac{1}{\sqrt{n}}\sum_{j\neq r+1}(y_j-m))}\\
&& = \frac{n!}{(r!)^2\sqrt{n}}f(\mu+\frac{h}{\sqrt{n}}) 
\int \frac{dt}{2\pi} e^{it(l-\frac{h}{n}+\frac{m-\mu}{\sqrt{n}})}
\left (\int_{-\infty}^{\mu+\frac{h}{\sqrt{n}}}dy f(y) e^{it\frac{m-y}{\sqrt{n}}}\right )^r
\left (\int^{\infty}_{\mu+\frac{h}{\sqrt{n}}}dy f(y) e^{it\frac{m-y}{\sqrt{n}}}\right 
)^r 
\eeq 
The integrals between parentheses need only be calculated to order $\frac{1}{n}$ to find the
limit when $r =\frac{n-1}{2} \rightarrow \infty$ \\
We do this as follows:
\beq \int_{-\infty}^{\mu + \frac{h}{\sqrt{n}}}dyf(y)e^{it\frac{m-y}{\sqrt{n}}} & =  
\displaystyle \int_{-\infty}^{\mu}dyf(y)e^{it\frac{m-y}{\sqrt{n}}} + \int_{\mu}^{\mu + 
\frac{h}{\sqrt{n}}} 
dyf(y)e^{it\frac{m-y}{\sqrt{n}}} & = \displaystyle I_1(t) + I_2(t) \eeq

Because $f$ is integrable and has moments of order 1 and 2, the first two derivatives of
the function $I_1(t)$ can be calculated by differentiating under the integral sign.
This allows us to write down the following expansion:
$$ \displaystyle I_1(t) = \frac{1}{2} +i\frac{t}{\sqrt{n}}\int_{-\infty}^{\mu}dy f(y)(m-y) - \frac{t^2}{2n} 
\int_{-\infty}^{\mu}dy f(y)(y-m)^2 + o(\frac{t^2}{n})$$ or
$$ \displaystyle I_1(t) = \frac{1}{2} +i\frac{t}{\sqrt{n}}\kappa -\frac{t^2}{2n}\sigma^2_l +o(\frac{t^2}{n})$$

where we have used the definitions:~$\kappa = \int_{-\infty}^{\mu}dy f(y)(m-y) = \frac{m}{2} - \int_{-\infty}^{\mu}dy 
f(y)y$ and $\sigma^2_l = \int_{-\infty}^{\mu}dy
f(y)(y-m)^2$\\

The second $y$-integral in $g(l,h)$ can be written:
$$\int^{\infty}_{\mu + \frac{h}{\sqrt{n}}}dyf(y)e^{it\frac{m-y}{\sqrt{n}}} = I_1'(t) - I_2(t)$$
with $$I'_1(t) = \int^{\infty}_{\mu} dy f(y)e^{it\frac{m-y}{\sqrt{n}}} = \frac{1}{2} - 
i\frac{t}{\sqrt{n}}\kappa -\frac{t^2}{2n}\sigma^2_r + o(\frac{t^2}{n}) $$
and an obvious definition for $\sigma_r$\\
 
From these expressions, it is clear that in order to calculate the product $(I_1+I_2)(I'_1-I_2)$ to order $\frac{1}{n}$ 
we only need $I_2$ to order $\frac{1}{\sqrt{n}}$, which is immediate: $\displaystyle I_2 =
\frac{hf(\mu)}{\sqrt{n}} + \frac{1}{n}I'_2$\\

We therefore get the following expression for the two $y$ integrals in $g$:\\

$\displaystyle I_1(t)+I_2(t) = \frac{1}{2} +  \frac{h}{\sqrt{n}}f(\mu) + i\frac{t}{\sqrt{n}}\kappa 
+ \frac{1}{n}I'_2 -\frac{t^2}{2n}\sigma^2_l$\\

and\\

$\displaystyle I'_1(t)-I_2(t) = \frac{1}{2} -  \frac{h}{\sqrt{n}}f(\mu) - i\frac{t}{\sqrt{n}}\kappa
- \frac{1}{n}I'_2 -\frac{t^2}{2n}\sigma^2_r$\\

The $\frac{1}{\sqrt{n}}$ and the $I'_2$ terms disappear in the product which, to order $\frac{1}{n}$ reads:

\beq \int_{-\infty}^{\mu + 
\frac{h}{\sqrt{n}}}dyf(y)e^{it\frac{m-y}{\sqrt{n}}}\int^{\infty}_{\mu 
+ \frac{h}{\sqrt{n}}}dyf(y)e^{it\frac{m-y}{\sqrt{n}}} = & 
 \frac{1}{4}\left ( 1 + \frac{4 t^2\kappa^2-t^2\sigma^2 -4h^2 f(\mu)^2 - 8 i 
t \kappa h f(\mu)}{n} \right )&\eeq

The term between parentheses is an asymptotic expansion valid for small $\frac{t}{\sqrt{n}}$. However, when
raised to the power $r = \frac{n-1}{2}$, missing terms of higher order in $\frac{1}{n}$ give no contributions 
in the $n \rightarrow \infty$ limit which is thus valid for any $t$ and reads: 
$$\displaystyle e^{2t^2\kappa^2-\frac{t^2\sigma^2}{2} -4 i t \kappa h 
f(\mu)} e^{-2h^2 f(\mu)^2}$$ 
The first exponential (with the sign of $t$ reversed) is recognized as the Fourier transform 
of a gaussian distribution of variance $\sigma^2 - 4 \kappa^2$ and expectation value $4\kappa h f(\mu)$\\
Therefore, changing $t$ to $-t$ (to perform the inverse transform) and neglecting terms with
negative powers of $n$ in the argument of the exponential yields after integration over $t$~:
$$\frac{1}{\sqrt{2\pi (\sigma^2 - 4 \kappa^2)}}e^{-\frac{(l-4\kappa h f(\mu))^2}{2(\sigma^2 - 
4 \kappa^2)}} e^{-2h^2 f(\mu)^2}$$
On the other hand, use of Stirling's formula to treat the numerical coefficient \\
$\frac{n!}{(r!)^2\sqrt{n}}f(\mu+\frac{h}{\sqrt{n}})\frac{1}{4^r}$ leads to 
$\sqrt{\frac{2}{\pi}}f(\mu)$ which checks the normalisation of our asymptotic joint
distribution.
\subsection{Result}
Overall, we have found that $$g(l,h) = \frac{2f(\mu)}{2\pi\sqrt{\sigma^2 - 4 
\kappa^2}}e^{-\frac{(l-4\kappa h f(\mu))^2}{2(\sigma^2 -
4 \kappa^2)}} e^{-2h^2 f(\mu)^2}$$
which shows that the two random variables $H$ and $L-4\kappa f(\mu) H$ are gaussian 
distributed and independnt with zero expectation value and variances $\frac{1}{4 f(\mu)^2}$ 
and $\sigma^2 -4 \kappa^2$ respectively.\\

From here one finds $V[L] = \sigma^2$ as expected, and $Cov[H,L] = 
\frac{\kappa}{f(\mu)}$, a possibly new result.\\

For a finite but large sample, $\bar{X}$ and $M$ are approximately normal
with E.V.'s $m$ and $\mu$ and covariance matrix:
\begin{displaymath}
\frac{1}{n}
\begin{pmatrix}
\;\sigma^2 \hfill \;\;\;\frac{\kappa}{f(\mu)} \hfill \\
\frac{\kappa}{f(\mu)} \hfill \;\;\;\;\frac{1}{4f(\mu)^2} \hfill
\end{pmatrix} 
\end{displaymath}
\section{Conclusions}
There are two major reasons which explain the predominance of the expectation value and the mean in the
definition and estimation of a central value for a probability distribution. The first is the linearity
of the barycentric processes, the second is the asymptotic ubiquitness of the Laplace-Gauss law for which
the expectation value is well defined, the sample mean being the 'best' (in various ways) estimator thereof.
However, the sample mean is of no help for the estimation of the central value of more dispersed distributions 
like Cauchy's and can be seriously flawed by the contamination of a standard sample by outliers. In that case,
the median which is by construction far less sensitive to outliers can be a better choice. It has several
fine qualities but also several drawbacks which have been reviewed to some extent. However, the often posed
problem of the calculation of the median moments to estimate its average value or its uncertainty have been
shown to be answers to the wrong questions. Moments and quantiles are two different ways of trying to sum up the 
complexity of a full probability distribution with a few numbers. The expectation value and the variance belong
to the first approach, whilst the median belongs to the second. It is therefore clear that the sample median 
must used for estimating the distribution median without reference to expectation values and that interval
estimation of this median is most simply done by using quantiles or approximating intervals of given probability
content for the sample median through the use of the dispersion measures which have been suggested above.
\section{Appendix}
\begin{center} {\bf \small On the impossibility of an unbiassed ratio estimator when both terms have errors.} 
\end{center}
\scriptsize
The problem of building an unbiassed estimator for a ratio amounts to finding a function $f$ of two 
variables such that \be E[f(X,Y)] = \frac{E[Y]}{E[X]} \label{eqf} \ee for any two independant random variables $X$ and 
$Y$\\

Since $X$ and $Y$ can be taken as having fixed (arbitrary) values $x_0$ and $y_0$ (in which case their densities
are Dirac's $\delta$'s), eq (\ref{eqf}) implies that $$f(x,y) = y/x \;\;\forall (x,y) \in {\Bbb R}^* \times {\Bbb R}$$

However, this would impose $E[Y/X] = E[Y]/E[X]$ hence $E[1/X] = 1/E[X]$ for any random variable $X$ for which both sides 
make sense, and this is well known to be false as can be shown by elementary examples. \\

On the other hand, what is required of $E[1/X]$ corresponds to what we have called the invariance property of the median 
($M[1/X] =1/M[X]$) and one could therefore hope to solve the problem using this statistics rather than the mean.

But by the same reasoning used for the mean, solving $M[f(Y,X)] = M[Y]/M[X]$ leads to $f(x,y) = y/x$.
However, it is not true that $M[Y/X] = M[Y]/M[X]$ in general; for example, the ratio of a $\gamma(2,1)$ and a 
$\gamma(1,1)$ independant random variables has median $1.+\sqrt{2.} = 2.414$ instead of the 
expected $\approx 2.421$ which is the ratio of the medians. 

Therefore the median does not solve the problem of a distribution free, strictly unbiassed estimator of a ratio in the 
above sense.


\begin{thebibliography}{5}
\bibitem{vocabl} The same term, median, is used for the probability distribution central value 
and for the sample central value when central is thought of as 'splitting into equal halves', whereas there are two 
terms, mean (or average) for the sample and expectation value 
for the distribution when central is understood as 'barycentric'. We shall therefore consistently 
use the expressions 'distribution median' and 'sample median' to avoid confusion.
\bibitem{beta} $B(k,l)$ stands for $\frac{\Gamma(k)\Gamma(l)}{\Gamma(k+l)}$. Therefore $F(Y_k)$ follows
a $\beta_1(k,n-k+1)$ probability distribution the c.d.f. of which is the corresponding normalized
incomplete $\beta$ integral.
\bibitem{note1}In principle, it is possible to have $F(x) =1/2$ for a
whole range of values of $x$. This means that the p.d.f. $f(x) = 0$ in that range, which is usually  
very unrealistic in physical problems, even though analytical examples are easy to build.
\bibitem{oldfoot}As a counter-example, it is enough to calculate
numerical approximations for $\chi^2_n$ distribution medians. For $n=1,2,3$ one finds respectively:
$.45 ; 1.38 ; 2.36$ to better than $.01$; however, $2*.45 \neq 1.38$ and $.45+1.38 \neq 2.36$
\bibitem{noreg} The problem is entirely different from the usual least square fit wherein the input
variables are assumed to be perfectly known.
\bibitem{jml} The constrained fit is equivalent to a least-square fit wherein weights (inverse variances) are 
calculated as functions of the parameter to be fitted.
\bibitem{note2}The $68\%$ choice is made by analogy with the probability content of a 
one-$\sigma$ half-width interval for a gaussian distributed random variable.
\end{thebibliography}
\end{document}